\documentclass[letterpaper,12pt]{article}

\usepackage[T1]{fontenc}
\usepackage{lmodern}          
\usepackage{microtype}        
\usepackage{setspace}
\usepackage[hidelinks]{hyperref}

\usepackage[english]{babel}
\usepackage[utf8x]{inputenc}
\usepackage{amsmath,amsthm}
\usepackage{algorithm}
\usepackage[noend]{algpseudocode}
\usepackage{graphicx}
\usepackage[colorinlistoftodos]{todonotes}
\usepackage[margin=1in, letterpaper]{geometry}
\usepackage{amssymb}
\usepackage{float}
\usepackage{caption}

\newcommand{\kw}[1]{\textbf{#1}}

\newcommand{\algname}[1]{\textsf{#1}}

\DeclareMathOperator{\degop}{deg}

\newcommand{\algcomment}[1]{\textit{\textcolor{gray}{#1}}}

\usepackage[
backend=biber,
sorting=ynt
]{biblatex}

\title{Parameterized Algorithms for Spanning Tree \\
       Isomorphism by Redundant Set Size}
\author{%
  Fangjian Shen\thanks{Sun Yat-sen University. shenfj3@mail2.sysu.edu.cn
}  \ \ \  Yicheng Zheng\thanks{1931195@qq.com.  This work was done in part while the author was studying in Central South University} \ \ \ Wushao Wen\thanks{Sun Yat-sen University. wenwsh@mail.sysu.edu.cn} \ \ \  Hankui Zhuo\thanks{Nanjing University. hankz@nju.edu.cn}
}
\date{}

\begin{document}
\maketitle

\begin{abstract}
In this paper, we present fixed-parameter tractability algorithms for both the undirected and directed versions of the Spanning Tree Isomorphism Problem, parameterized by the size $k$ of a redundant set. A redundant set is a collection of edges whose removal transforms the graph into a spanning tree. For the undirected version, our algorithm achieves a time complexity of $O(n^2 \log n \cdot 2^{k \log k})$. For the directed version, we propose a more efficient algorithm with a time complexity of $O(n^2 \cdot 2^{4k-3})$, where $n$ is the number of vertices.
\end{abstract}

\thispagestyle{empty}  
\clearpage             
\setcounter{page}{1}   

\section{Introduction}

The Spanning Tree Isomorphism Problem (STIP) asks whether a given graph $G$ contains a spanning tree isomorphic to a target tree $T$. Formally, given a graph $G=(V_G,E_G)$ with $|V_G|=n$ vertices and a target tree $T=(V_T,E_T)$ on the same number of vertices, STIP is the problem of determining if there exists a bijection $\phi: V_G \rightarrow V_T$ and an edge subset $E_H \subseteq E_G$ with $|E_H| = n-1$, such that for any pair of vertices $u,v \in V_G$, $(u,v) \in E_H$ if and only if $(\phi(u), \phi(v)) \in E_T$. This problem is a classic in computational complexity, formally established as NP-complete by Garey and Johnson \cite{10.5555/574848}, with its structural properties first investigated in the seminal work of Papadimitriou and Yannakakis \cite{papadimitriou1982complexity}. Consequently, much of the subsequent research has focused on constrained variants of the problem. A notable example is the work by Cai and Corneil \cite{cai1995isomorphic}, which requires the solution to be a tree t-spanner—--a spanning tree where the distance between the endpoints of any edge from the original graph $G$ is at most $t$. However, their algorithm is applicable only to the case of $t=2$, as the problem becomes intractable for $t \geq 3$. While valuable for specific applications, this constraint is highly restrictive, rendering the approach inapplicable to the general STIP. Since STIP can be viewed as a special case of the Subgraph Isomorphism problem where the pattern is a tree and the host is a general graph, it naturally connects to two broad areas of research:


\textbf{Graph Isomorphism Problem.} The Graph Isomorphism Problem (GIP) seeks to determine if two given graphs of order $n$ are isomorphic. A key breakthough is provided by Babai \cite{babai2016graph}, who presented a quasi-polynomial time algorithm with complexity $2^{O((\log n)^c)}$ for a fixed constant $c$. However, the precise complexity of GIP remains unresolved; it is one of the few problems in NP not known to be in P or NP-complete. The absence of a general polynomial-time solution has motivated extensive research on GIP for restricted graph classes \cite{aho1974design, babai1982isomorphism, bodlaender1990polynomial, das2020polynomial, filotti1980polynomial, grohe2015isomorphism, hopcroft1974linear, miller1983isomorphism, neuen2016graph, juttner2018vf2++}. A notable early success in this direction is the linear-time algorithm for the Tree Isomorphism Problem (TIP) by Aho et al. \cite{aho1974design}. More recently, this focus has led to the development of fixed-parameter tractability (FPT) algorithms. For instance, Kratsch et al. \cite{kratsch2010isomorphism}proposed an FPT algorithm parameterized by the feedback vertex set size $k$, with a runtime of $O((2k + 4k \log k)^k \cdot kn^2)$ . For the parameter treewidth $k$, Lokshtanov et al. \cite{lokshtanov2017fixed} gave an algorithm with complexity $2^{O(k^2 \log k)} \cdot n^{O(1)}$, which was subsequently improved by Grohe \cite{grohe2020improved} et al. to $2^{O(k(\log k)^c)} \cdot n^3$. In addition, several studies have proposed FPT algorithms for GIP using other graph properties as parameters \cite{arvind2017finding, arvind2022testing, bulian2016graph, ccaugirici2019efficient, lokshtanov2022fixed, neuen2024isomorphism}.

\textbf{The Subgraph Isomorphism Problem.} The problem of determining whether a pattern graph $P$ is isomorphic to a subgraph of a host graph $G$ is NP-complete in its general form. Among seminal works \cite{ullmann1976algorithm, cordella2004sub, bonnici2013subgraph, kotthoff2016portfolios, asiler2022hygraph, micale2021temporalri}, the most influential is Ullmann's \cite{ullmann1976algorithm} recursive backtracking algorithm, which systematically explores all possible node mappings and has a worst-case complexity of $O(n^2 \cdot n!)$. Subsequent research has largely built upon Ullmann's approach. For instance, Cordella et al. \cite{cordella2001improved} introduced more effective pruning rules and an optimized node matching order, resulting in the VF2 algorithm, which is significantly more efficient on large-scale graphs. Carletti et al. \cite{carletti2017introducing} further optimized VF2 using heuristics, yielding the VF3 algorithm and substantial performance improvements in practice. Significant research has also addressed the problem through parameterized complexity, producing several FPT algorithms. A landmark result by Alon et al. \cite{alon1995color} introduced the color-coding technique, providing a randomized $2^{O(k)} \cdot n^{O(1)}$-time algorithm where $k = |V(P)|$ is the parameter. This established the first proof that SIP is FPT for this fundamental parameter. Further studies exploit structural graph properties to achieve greater efficiency. Courcelle's \cite{courcelle2000linear} Theorem implies that if the host graph $G$ has bounded treewidth $\text{tw}(G)$, a dynamic programming approach solves the problem in $k^{O(\text{tw}(G))} \cdot n^{O(1)}$ time. While parameterizing solely by the pattern's treewidth $\text{tw}(P)$ is W[1]-hard, combining it with $k$ enables advanced algorithms that use techniques like important separators to achieve runtimes of $2^{O(k log k)} \cdot n^{O(1)}$. Marx et al. \cite{marx2013everything} showed that if both $P$ and $G$ are planar, the complexity reduces to $2^{O(\sqrt{k})} \cdot n^{O(1)}$. Indeed, numerous studies have developed FPT algorithms for SIP using other graph properties as parameters \cite{marx2013everything}.

Although the GIP and SIP have been extensively studied over the past three decades, algorithms for GIP are not directly applicable to the STIP. While algorithms for SIP can theoretically be applied, their performance in the STIP is generally suboptimal. Classical SIP backtracking methods incur prohibitive $O(n!)$ complexity. Similarly, FPT algorithms which are parameterized by properties such as pattern size $k$ , fail fundamentally for the STIP, thereby negating the utility of the parameterization. STIP, however, possesses considerable practical value with wide-ranging applications. For instance, in communication networks and distributed systems \cite{bertalanivc2023graph, tran2008graph}, $G$ might represent a physical router network, while the target tree $T$ models a logical topology optimized for tasks—such as a small-diameter tree for rapid communication, a degree-constrained tree to prevent node overload during data aggregation, or a specific structure ensuring deadlock-free routing. In computational biology and cheminformatics \cite{merkys2023graph, willett1999matching}, STIP determines whether a known functional motif or structural backbone ($T$) exists within a larger molecular structure or network ($G$). In Very Large Scale Integration (VLSI) design \cite{kresh1987graph, mishra2017using}, it models embedding a tree-structured routing configuration into a chip's pathway grid. Consequently, STIP remains a significant research focus. However, as is known so far, there is no effective FPT algorithm for the general STIP problem.

\textbf{Contributions}: In this paper, we present two FPT algorithms for the STIP (one $O(n^2logn \cdot 2^{k\log k})$-time for the undirected version and another $O(n^2 \cdot 2^{4k-3})$-time for the directed version), where $k$ and $n$ are the size of the redundant set and the number of vertices in the input graphs, respectively. A redundant set in an undirected graph refers to a feedback edge set whose removal results a spanning tree. In a directed graph, a redundant set is an arc set such that, after removing it, the underlying graph of the remaining directed graph is a spanning tree. In other words, if a graph has $n$ vertices and the size of its redundant set is $k$, then this graph contains $n + k - 1$ edges (or arcs). To our best knowledge, we are the first to solve general STIP by FPT algorithm. Therefore, our work offers valuable insights that may inform the future development of parameterized algorithms for STIP.

The paper is organized as follows. In Section 2, we introduce the necessary preliminaries and notations and provide a formal definition of STIP. We then present our FPT algorithm for the undirected version of STIP in Section 3 and for the directed version in Section 4. Section 5 concludes.

\section{Preliminaries}

Let $G = (V_G, E_G)$ be an undirected graph, where $V_G$ is the set of vertices and $E_G$ is the set of edges. The set of neighbors of a vertex $v \in V_G$ is its neighborhood, denoted by $N_G(v)$. The number of edges incident on $v$ is its degree, denoted by $deg(v) = |N_G(v)|$. A path $P$ in $G$ is a sequence of vertices $(v_1, v_2, \dots, v_m)$ such that $(v_i, v_{i+1}) \in E_G$ for all $1 \le i < m$. For a subset of edges $E' \subseteq E_G$, we let $V(E')$ denote the set of all endpoints of the edges in $E'$. The graph obtained by removing an edge set $E'$ from $G$ is denoted by $G - E'$, which has vertex set $V_G$ and edge set $E_G \setminus E'$. An edge set $E' \subseteq E_G$ is called a feedback edge set of $G$ if the graph $G - E'$ is a tree. A subgraph of $G$ that is a tree and contains all vertices of $V_G$ is a \textit{spanning tree}. A \textit{depth-first search (DFS) tree} of $G$ is a spanning tree formed by a depth-first search starting from a given root. The sequence in which vertices are first visited during the traversal defines the \textit{DFS order}.

A directed graph (digraph) is denoted by $D = (V_D, A_D)$, where $A_D$ is a set of arcs. An arc is an ordered pair of vertices $(u, v)$, where $u$ is the \textit{tail} and $v$ is the \textit{head}. The \textit{underlying graph} of a digraph $D$ is the undirected graph obtained by replacing each arc $(u, v) \in A_D$ with an undirected edge $(u, v)$. An arc set $A' \subset A_D$ is a feedback arc set of $D$ if the underlying graph of $D - A'$ is a tree.

In an undirected graph $G$, two vertices $u, v \in V_G$ are \textit{connected} if there is a path between them. A graph is \textit{connected} if every pair of its vertices is connected. A \textit{connected component} of $G$ is a maximal connected subgraph. In a digraph $D$, a vertex $v$ is \textit{reachable} from a vertex $u$ if there is a path from $u$ to $v$. A digraph is \textit{strongly connected} if every vertex is reachable from every other vertex. It is \textit{weakly connected} if its underlying graph is connected. Analogously, a digraph can be partitioned into \textit{weakly connected components} and \textit{strongly connected components}, which are its maximal weakly and strongly connected subgraphs, respectively. We denote the connected component containing a vertex $v$ by $C(v)$.

In a connected undirected graph $G$, two vertices $u, v \in V_G$ are \textit{2-edge-connected} if they remain connected after the removal of any single edge. Similarly, they are \textit{2-vertex-connected} if they remain connected after the removal of any single vertex other than $u$ and $v$. A maximal 2-edge-connected subgraph of $G$ is called an \textit{edge-biconnected component} (or a bridge-block). A maximal 2-vertex-connected subgraph is a \textit{vertex-biconnected component} (often simply called a biconnected component or a block). We say that $C = (v_1,v_2,\cdots,v_k = v_1)$ is called a weak cycle in $D$ if $C$ is a cycle in $G$.

A graph $G$ that is a tree is called an \textit{unrooted tree} if no specific vertex is designated as the root. If a vertex $r$ is designated as the root, it is a \textit{rooted tree}. In a rooted tree $T$ with root $r$, for any two vertices $u, v \in V(T)$, $u$ is an \textit{ancestor} of $v$ if $u$ lies on the unique path from $r$ to $v$. For a vertex $v \in V_T$, its \textit{depth} is the number of edges on the path from the root to $v$. The \textit{height} of a tree is the maximum depth among all its vertices.

Two rooted trees, $T_1 = (V_1, E_1, r_1)$ and $T_2 = (V_2, E_2, r_2)$, are \textit{isomorphic}, denoted $T_1 \cong T_2$, if there exists a bijection $\phi : V_1 \to V_2$ such that $\phi(r_1) = r_2$ and for any pair of vertices $u, v \in V_1$, $(u, v) \in E_1$ if and only if $(\phi(u), \phi(v)) \in E_2$. Similarly, two graphs $G_1 = (V_1, E_1)$ and $G_2 = (V_2, E_2)$ are \textit{isomorphic}, denoted $G_1 \cong G_2$, if such a bijection exists that preserves adjacency. A graph where each vertex is assigned a label is a \textit{labeled graph}. For two labeled graphs $G_1$ and $G_2$, an isomorphism $\phi$ must also preserve labels, meaning that for any vertex $v \in V_1$, the label of $v$ must be identical to the label of $\phi(v)$.

The definitions of the problems considered in this paper are given as follows.\newline

\noindent
\textbf{Parameterized Undirected Spanning Tree Isomorphism Problem (PUSTIP)}

\noindent
Input: An undirected graph $G = (V_G, E_G)$ and a tree $T = (V_T, E_T)$, where $|V_G| = |V_T| = n$ and $|E_G| = m$.

\noindent
Parameter: the size $k = m - (n - 1)$ of the redundant set of $G$.

\noindent
Output: return YES if $\exists \ H \subset G, \ H \cong T$; otherwise NO.\newline

\noindent
\textbf{Parameterized Directed Spanning Tree Isomorphism Problem (PDSTIP)}

\noindent
Input: A directed graph $D = (V_D, A_D)$ and a tree $T = (V_T, E_T)$, where $|V_D| = |V_T| = n$ and $|A_D| = m$.

\noindent
Parameter: the size $k = m - (n - 1)$ of the redundant set of $D$.

\noindent
Output: return YES if $\exists \ F \subset D, \ F \cong T$; otherwise NO.

\section{Algorithm for the PUSTIP}

Our approach to solving PUSTIP is inspired by the AHU and Ullmann algorithms, reframing the isomorphism detection problem as a two-phase matching process. Specifically, by analyzing the structural differences between the input graph and the target tree, we propose an efficient algorithm that isolates a \textit{topological kernel} within the graph. This kernel, whose size is bounded by a function of the parameter $k$, is then leveraged to guide a recursive search for a valid isomorphism.

To achieve this, we first need to address the case where the input graph $G$ contains exactly one cycle ($k = 1$). Here, the problem admits a polynomial-time solution: we enumerate all cycle edges, remove each in turn to generate a spanning tree, and verify isomorphism with the target tree $T$ using the AHU algorithm. In what follows, we assume that $G$ contains at least two cycles. \newline

\noindent\textbf{Definition 3.1.}
Let $G = (V_G, E_G)$ be a graph and $T = (V_T, E_T)$ be a tree. We write $G \cong T$ to denote that $G$ contains a spanning tree $H$ that is isomorphic to $T$.\newline

\noindent
\textbf{Lemma 3.1.}
Let $G = (V_G, E_G)$ be a graph with $|E_G| = |V_G| + k - 1$. For any vertex $v \in V_G$, its neighborhood $N_G(v)$ contains at most $2k$ vertices $u$ with the property that in the subgraph $G' = G - \{(u,v)\}$, the connected component containing $u$, denoted $C_u$, either contains a cycle or includes the vertex $v$.
\begin{proof}
A graph $G$ satisfying the given edge constraint can be decomposed into a spanning tree and a set of $k$ redundant edges $G = T' \cup \{e_1, \dots, e_k\}$, $|E(T')| = |V_G| - 1$, $|E_G \setminus E_{T'}| = k$. Let $T'$ be the Breadth-First Search (BFS) spanning tree of $G$ rooted at a vertex $v \in V_G$. Consequently, any edge in $E_G \setminus E(T')$ cannot be incident to the root $v$. Then the set of children of $v$ in $T'$ is denoted as $S_v$. Note that $N_G(v) = S_v$. 

We define a vertex $u \in S_v$ as \textit{invalid} if, in the graph $G' = G - \{(u,v)\}$, the connected component $C_u$ containing $u$ is not a tree or if $v \in V_{C_u}$. Otherwise, $u$ is \textit{valid}. Initially, considering only the tree $T'$, every child $u \in S_v$ is valid, as removing the edge $(u,v)$ simply disconnects the subtree rooted at $u$ from the rest of $T'$. We now analyze the effect of adding the $k$ edges from $E_G \setminus E(T')$ one by one. Let $I \subseteq S_v$ be the set of invalid children at any step. When we add an edge $e = (x,y)$, two cases arise:

\noindent\textbf{Case 1: Both $x$ and $y$ are in the same subtree $T_u$ for some $u \in S_v$.}
Adding $e$ creates a cycle within the subtree $T_u$. This makes the child $u$ invalid, because after removing $(u,v)$, the component $C_u$ now contains a cycle. If $u \notin I$, then $u$ becomes a new illegal child node, $|I| = |I| + 1$. If $u \in I$, then $I$ does not change. Thus, increasing $|I|$ by at most 1.

\noindent\textbf{Case 2: $x$ and $y$ are in different subtrees, $T_a$ and $T_b$, where $a, b \in S_v$ and $a \neq b$.}
Adding the edge $e = (x,y)$ connects the two subtrees. Consider $a \not \in I$. After removing $(a,v)$, $a$ can still reach $v$ via the path $a \leadsto x \to y \leadsto b \to v$. Thus, $a$ becomes invalid. Similarly, $b$ becomes invalid. If both $a$ and $b$ were previously valid, $|I|$ increases by 2. If exactly one of them was already invalid, $|I|$ increases by 1. If both were already invalid, $|I|$ remains unchanged. Therefore, adding the edge $e$ makes at most 2 new vertices in $S_v$ invalid. Since only a total of $k$ edges can be added, $|I| \leq 2k$.
\end{proof}

\noindent\textbf{Lemma 3.2.}
Let $G$ be a graph, $T_1$ and $T_2$ be trees. If $G \cong T_1$, and $T_1 \cong T_2$, then $G \cong T_2$.
\begin{proof}
By the definition of graph isomorphism, these two isomorphisms imply the existence of two bijections $\phi: V_G \to V_{T_1}$ and $\psi: V_{T_1} \to V_{T_2}$, simultaneously satisfying the property:
\begin{flalign*}
    &1.\ \forall u, v \in V_G,\; (u, v) \in E_G \iff (\phi(u), \phi(v)) \in E_{T_1} & \\
    &2.\ \forall x, y \in V_{T_1},\; (x, y) \in E_{T_1} \iff (\psi(x), \psi(y)) \in E_{T_2} &
\end{flalign*}
Consider the composite function $\theta = \psi \circ \phi$. Since $\phi$ and $\psi$ are bijections, their composition $\theta$ is also a bijection $V_G \to V_{T_2}$, that is: $\theta(u) = \psi(\phi(u)), \forall u \in V_G$. So $\forall u, v \in V_G$, we have:
\begin{align*}
(u,v) \in E_G \iff (\phi(u), \phi(v)) \in E_{T_1} \iff (\psi(\phi(u)), \psi(\phi(v))) \in E_{T_2} \iff (\theta(u), \theta(v)) \in E_{T_2}
\end{align*}
That is, the spanning tree isomorphism has transitivity.
\end{proof}

\noindent\textbf{Definition 3.2.} A graph is \textit{reducible} if it contains at least one vertex of degree less than 3. A non-empty graph that is not reducible is called a \textit{graph kernel}. A graph kernel, therefore, is a graph where every vertex has a degree of at least 3.\newline

To isolate the topologically complex part of an input graph $G$, we define a reduction procedure that systematically transforms $G$ into its graph kernel. This process iteratively removes vertices with simple local structures. It consists of two repeatedly applied operations until no such vertices remain:

\begin{enumerate}
    \item \textbf{Leaf Trimming}: If a vertex $v$ has a degree of 1, remove $v$ and its incident edge.
    \item \textbf{Path Contraction}: If a vertex $v$ has a degree of 2 with neighbors $u$ and $w$, remove $v$ and its two incident edges, $(u,v)$ and $(v,w)$, and add a new edge $(u,w)$.
\end{enumerate}

This procedure continues until the graph has no vertices of degree 1 or 2, resulting in the graph kernel (as illustrated conceptually in Figure \ref{tab:make-contractible}). We now prove that the size of this kernel is bounded by the parameter $k$.\newline

\noindent\textbf{Lemma 3.3.}
Let $G=(V_G,E_G)$ be a graph such that $|E_G|=|V_G|+k-1$. Let $G'=(V_G',E_G')$ be the graph kernel obtained by exhaustively applying the leaf trimming and path contraction operations to $G$. If $G'$ is non-empty, then $|V_G'| \leq 2k-2$.

\begin{proof}
First, we observe that both reduction operations preserve the value of $|E_G| - |V_G|$. A leaf trimming operation removes one vertex and one edge, so $|E_G|-|V_G|$ is unchanged. A path contraction removes one vertex and two edges, then adds one edge, resulting in a net change of $-1$ vertex and $-1$ edge. Thus, the condition $|E'| = |V'| + k - 1$ remains valid for $G'$. By Definition 3.2, every vertex in the non-empty kernel $G'$ must have a degree of at least 3. Let $D'_{G}$ be the sum of degrees of all vertices in $G'$. We have a lower bound based on this property:
\begin{equation}\label{Lemma 341}
D'_{G} = \sum_{v \in V'} \text{deg}(v) \geq 3|V'|
\end{equation}
By the handshaking lemma, the sum of degrees is equal to twice the number of edges $D'_{G} = 2|E'|$. Substituting the invariant relationship $|E'| = |V'| + k - 1$ into this equation gives:
\begin{equation}\label{Lemma 342}
D'_{G} = 2(|V'| + k - 1) = 2|V'| + 2k - 2
\end{equation}
By combining the lower bound for $D'_{G}$ with this equality, we obtain:
\begin{equation}\label{Lemma 343}
3|V'| \leq 2|V'| + 2k - 2
\end{equation}
By \eqref{Lemma 341} \eqref{Lemma 342} \eqref{Lemma 343} we have $|V'| \leq 2k - 2$. This shows that the number of vertices in the graph kernel is bounded by $2k-2$.
\end{proof}
\noindent
\begin{figure}[ht]
\centering
\fbox{
\begin{minipage}{\dimexpr\textwidth-2\fboxsep-2\fboxrule\relax}
\linespread{1}\selectfont
\vspace{2mm}

\begin{tabular}{@{}l}
\kw{procedure} \algname{Make-Contractible}($G$) \\[6pt]

\kw{while} $G$ is contractible \kw{do} \\
    \quad \kw{if} there exists a vertex $v \in V(G)$ with $\degop_G(v) = 1$ \kw{then} \\
    \quad\quad \algcomment{Perform Leaf Pruning} \\
    \quad\quad $\star$ Let $u$ be the unique neighbor of $v$. \\
    \quad\quad $G \leftarrow G - \{v, (u,v)\}$. \\
    \quad \kw{else} \\
    \quad\quad $\star$ Let $v$ be a vertex in $G$ with $\degop_G(v) = 2$. \\
    \quad\quad \algcomment{Perform Path Suppression} \\
    \quad\quad $\star$ Let $u$ and $w$ be the two neighbors of $v$. \\
    \quad\quad $G \leftarrow G - \{v, (u,v), (v,w)\}$. \\
    \quad\quad \kw{if} $u \neq w$ and $(u,w) \notin E(G)$ \kw{then} \\
    \quad\quad\quad $G \leftarrow G \cup \{(u,w)\}$. \\
    \quad\quad \kw{end if} \\
    \quad \kw{end if} \\
\kw{end while} \\[6pt]

\kw{return} $G$ \\[6pt]

\kw{end procedure} \\
\end{tabular}
\vspace{2mm}
\end{minipage}
}
\caption{Make Contractible Procedure} 
\label{tab:make-contractible} 
\end{figure}

\noindent\textbf{Definition 3.3.}
Let $G = (V_G, E_G)$ be a graph. For a vertex $v \in V_G$ and one of its neighbors $u \in N_G(v)$, consider the connected component of the graph $G-\{v\}$ that contains $u$, denoted by $C(u)$. The vertex $u$ is defined as a \textit{tree-like neighbor} of $v$ if the component $C(u)$ is a tree and $V_{C(u)} \cap N_G(v) = \{u\}$, otherwise, called a \textit{non-tree-like neighbor}.\newline

\noindent\textbf{Definition 3.4.} In an undirected graph $G=(V_G,E_G)$ with a graph kernel $G'=(V_G',E_G')$, a vertex $v \in V_G$ is defined as an \textit{anchor point} if it corresponds to a vertex $v' \in V_G'$ under a mapping $\delta$, such that $\delta(v')=v$.\newline

\noindent\textbf{Definition 3.5.} In an undirected graph $G=(V_G,E_G)$ with a graph kernel $G'=(V_G',E_G')$, a neighbor anchor set of vertex $v \in V_G$ is defined as the first anchor points encountered when performing DFS starting from each of its neighbors.\newline

The overall algorithm is outlined as follows. The process begins with two preprocessing steps: first, we perform a single depth-first search on the target tree $T$ to establish a canonical DFS traversal order of its vertices. Second, we apply the \textit{Make-Contractible} procedure above to the input graph $G$, reducing it to its topological core, the graph kernel $G'$.

The main algorithm then proceeds by iterating through each vertex $v \in V_G$ as a potential root of the spanning tree. For each such potential root, the algorithm further iterates through all possible permutations $\pi$ of the anchor points in $G'$. Each permutation $\pi$ represents a \textbf{hypothesized} DFS traversal order of these anchor points within a potential spanning tree rooted at $v$. For each $(v, \pi)$ pair, a recursive matching procedure is initiated. This procedure performs a coupled, depth first search of $G$ and $T$ starting from an initial mapping of a subgraph root $r_G$ to a subtree root $r_T$. The matching logic at each recursive step is divided into two phases:

\noindent\textbf{Phase 1: Tree-like Components Deterministic Matching}
The algorithm first prunes the search space by matching the tree-like neighbors of $r_G$. For each such neighbor, the algorithm computes the isomorphism type of its attached component. It then searches for an unmatched child of $r_T$ that roots a subtree with identical isomorphism type. If a corresponding subtree is found, the algorithm establish mapping and mark this child as matched. If for any tree-like neighbor of $r_G$, no such unmatched subtree can be found, the current permutation is proven invalid, and the algorithm returns. If they do match, a mapping is established for all tree-like neighbors. This partial isomorphism is recorded, and the corresponding nodes are removed from consideration in the next phase.

\noindent\textbf{Phase 2: Non-tree-like Components Recursive Search}
After resolving the tree-like components, the algorithm confronts the more complex problem of matching the remaining non-tree-like neighbors of $r_G$ to the remaining children of $r_T$. Let $x$ be the number of remaining neighbors of $r_G$ and $y$ be the number of remaining children of $r_T$.
\begin{itemize}
    \item If $x < y$, a valid mapping is impossible, the algorithm returns and verifies the next permutation.
    \item If $x \geq y$, we must select $y$ neighbors from the $x$ available to map to the children of $r_T$, which means we need to disconnect $x - y$ edges connected to non-tree-like neighbors, and the disconnection of specific edges is guided by $\pi$.
\end{itemize}
The guidance mechanism works as follows: the permutation $\pi$ defines the required sequence in which anchor points must be visited in a valid DFS traversal. The algorithm performs a DFS step from $r_G$ to identify the next anchor point it would encounter. It then verifies if this discovered anchor point is indeed the next one expected according to the order in $\pi$. If the discovered order is inconsistent with $\pi$, it signifies that the current hypothesis $\pi$ is incorrect for this search branch. 

The recursion proceeds by matching a valid neighbor pair and descending to the next level. A valid isomorphism is found when a complete mapping for all vertices is constructed and the algorithm terminates. If all potential roots and all permutations are exhausted without success, the algorithm correctly concludes that no such spanning tree exists. The pseudocode is provided in Algorithm 1.\newline

\begin{figure}[!t]
    \centering
    \begin{algorithm}[H]
      \footnotesize
      \caption{ALG-PUSTIP}
      \label{alg:directed_search_cn}
      \begin{algorithmic}[1]
        \Statex \textbf{Input:} An undirected graph $G = (V_G, E_G)$, an undirected tree $T = (V_T, E_T)$, and a parameter $k$
        \Statex \textbf{Output:} YES if a spanning tree of $G$ is isomorphic to $T$; and NO otherwise
        \Function{Main}{$G, T$}
          \State $t \gets \text{DFS}(\text{root of } T)$
          \State $G' \gets \text{Make-Contractible}(G)$
          \For{each vertex $v \in V_G$}
            \For{each permutation $\pi$ starting from $v$ in $G'$}
              \State $pos_G, \ pos_T \gets 1$
              \State $\text{Match} \gets \emptyset$
              \State $\text{Match(v)} \gets \text{root of T}$
              \If{\Call{search}{$v$, \text{root of } $T$, $\pi$, $pos_G$, $pos_T$}}
                \State \Return Yes
              \EndIf
            \EndFor
          \EndFor
          \State \Return NO
        \EndFunction
        
        \Function{search}{$r_G, r_T, \pi, pos_G, pos_T$}
          \For{each son $u$ of $r_G$ where $C(u)$ is a subtree}\Comment{Match subtree neighbors}
            \State $\text{found} \gets \textbf{false}$
            \For{each son $u'$ of $r_T$}
              \If{$C(u') \cong C(u)$ and $u'$ not matched in $\text{Match}$}
                \State $\text{Match}(u') \gets u$
                \State $\text{found} \gets \textbf{true}$
                \State \textbf{break}
              \EndIf
            \EndFor
            \If{not found}
              \State \Return \textbf{false}
            \EndIf
          \EndFor
          
          \State $x \gets \text{number of unmatched sons of } r_G$
          \State $y \gets \text{number of unmatched sons of } r_T$
          \If{$x < y$}
            \State \Return \textbf{false}
          \EndIf
          \If{$x = 0$ \textbf{and} $y = 0$} \Comment{Successful match on current branch}
            \State \Return \textbf{true}
          \EndIf
          
          \State $S \gets [\,]$
          \For{each unmatched son $u$ of $r_G$}
            \State $id \gets \text{fetch}(r_G, u, \pi)$\Comment{Fetch next anchor points}
            \State $S.\text{append}((u, id))$
          \EndFor
          \State \text{sort } $S$ \text{ by } $id$
          
          \For{each $(nxt, id)$ in $S$}
            \If{$id \neq pos_G + 1$} \Comment{Anchor point deviates from the order of $\pi$}
              \State \Return \textbf{false}
            \EndIf
            \State $dis \gets GetDistance(r_G, nxt)$
            \State $v' \gets t[pos_T + dis]$ \Comment{Next node in DFS order}
            \State $pos_G \gets pos_G + 1$
            \State $pos_T \gets pos_T + dis$
            \State $\text{Match}(v') \gets nxt$
            \If{not \Call{search}{$nxt, v', \pi, pos_G, pos_T$}}
              \State \Return \textbf{false}
            \EndIf
          \EndFor
          \State \Return \textbf{true}
        \EndFunction
      \end{algorithmic}
    \end{algorithm}
    \captionsetup{skip=0.25pt}
    \caption{The algorithm for the PUGIP}
    \label{fig:pugip_algorithm}
\end{figure}

\noindent\textbf{Theorem 3.1.} Define the subtree neighbor set of $v_G \in V_G$ as $S_1$, and the non-subtree neighbor set as $S_2$, the set of child nodes of $v_T \in V_T$ be $S_T$. Let $T_s(w)$ be the subtree of $T$ rooted at $w \in S_T$. If there exists an isomorphism mapping $\phi: V_G \rightarrow V_T$ satisfying $\phi(v_G) = v_T$. Then for all $u_1 \in S_1, u_2 \in S_2, w_1, w_2 \in S_T$, such that $w_1 = \phi(u_1), w_2 = \phi(u_2)$, swapping their matching order does not affect the algorithm's correctness. That is, matching $u_1$ first, then $u_2$, is equivalent in terms of solution space identification to matching $u_2$ first, then $u_1$, and will not lose any legal mappings, nor introduce any pseudo-mappings.
\begin{proof}
Assume, for the sake of contradiction, that exists a correct mapping $\phi: V_G \rightarrow V_T$: $\phi(v_T) = v_G, \phi(u_1) = w_1, \phi(u_2) = w_2$, that is, $C(u_1) \simeq T_s(w_1)$ and $C(u_2) \simeq T_s(w_2)$.
Let the original order be $\sigma_1=(\dots,u_2,u_1,\dots)$, and the swapped order be $\sigma_2=(\dots,u_1,u_2,\dots)$. Under $\sigma_2$, the algorithm cannot find the originally existing isomorphism mapping $\phi$, i.e., the solution is missed. Since we know a legal mapping $\phi$ exists satisfying: $\phi(v_G)=v_T, \phi(u_1)=w_1, \phi(u_2)=w_2, C(u_i) \cong T_s(w_i)$ for $i=1,2$, then the only possible reason for the matching to fail is: during the search process, $u_1$ was incorrectly mapped to some node other than $w_1$, causing $w_1$ to be subsequently unavailable for the mapping of $\phi(u_1)$.

Consider the child node set $S_T$ of the target tree's vertex $v_T$, which contains $w_1, w_2, \dots$. Since swapping the order only affects the matching order of $u_1$ and $u_2$, and other adjacent nodes (like $w_3, w_4$, etc.) did not successfully match with $u_1$ in the original order (otherwise the original $\phi$ would not exist), then after swapping the order, $u_1$ should not be preferentially matched to these nodes. Therefore, if $u_1$ fails to match its corresponding $w_1$, it must have been incorrectly matched to another candidate node. And this node must satisfy:
\begin{itemize}
    \item Not yet matched by another $u \in N_G(v_G)$;
    \item Isomorphic to the subgraph of $u_1$, $C(u_1)$;
    \item In $S_T$, and is attempted before $w_1$ in $\sigma_2$.
\end{itemize}

Combining with the fact that in the original matching order $\sigma_1$, $u_1$ was processed after $u_2$, and $w_2$ corresponds to $\phi(u_2)$, after swapping the order, $u_1$ is searched first, at which point $w_2$ is still unused, and thus might be mistakenly used to match $u_1$.

In summary, if the solution is missed, the only reasonable explanation is: $u_1 \to w_2 \implies C(u_1) \simeq T(w_2)$. Since for $\forall u_1 \in S_1$, by Lemma 3.2 we can get: $T(w_1) \simeq T(w_2)$, therefore $C(u_2) \simeq T(w_1)$, swapping the matching order does not cause solution loss, which means a valid isomorphism still exists when swapping matches. But this contradicts our initial assumption that the solution is missed. Therefore, the original assumption is false, and swapping matching orders does not affect correctness. 
\end{proof}

\noindent\textbf{Theorem 3.2.} Algorithm ALG-PUSTIP can solve the PUSTIP in runtime $O(n^2logn \cdot 2^{klogk})$.
\begin{proof} Given an instance of PUSTIP ($G$, $T$, $k$). ALG-PUSTIP first computes the DFS order of tree $T$ (Step 2) and generates the graph kernel $G'$ (Step 3) in linear time $O(n)$. It then iterates over each vertex $v \in V_G$ (Step 4), matching $v$ to the designated root of $T$. In Step 5, it enumerates all permutations of anchor points in $G'$. By Lemma 3.3, the total number of permutation is $(2k-2)!$ in the worst case. Note that:
$$
\begin{aligned}
k! \leq k^k \leq 2^{k \log k}
\end{aligned}
$$
Thus, the overall complexity is $O(n \cdot 2^{k \log k} \cdot T_{\text{search}})$, where $T_{\text{search}}$ is the cost of one $\text{search}$ call (Step 9). Steps 13 to step 21 perform deterministic tree-component matching in $O(n)$ time. Steps 29 to step 32 obtain the neighbor anchor set of $r_G$ and sort them by $\pi$, which costs $O(n \log n)$ time. In Steps 33 to step 42, the algorithm matches the remaining neighbor nodes recursively, which costs $O(n)$ in total. Thus:
$$
\begin{aligned}
T_{\text{search}} =O(n)+O(nlogn)+O(n)=O(nlogn)
\end{aligned}
$$
Let $T(n,k)$ denote the total complexity. The overall complexity is determined by the main loop and the search calls:
$$
\begin{aligned}
T(n,k) = O(n \cdot 2^{k \log k} \cdot T_{\text{search}})= O(n^2logn \cdot 2^{k \log k})
\end{aligned}
$$
\end{proof}


\section{Algorithm for the PDSTIP}

In this section, we shift our focus to the directed version of the problem: determining whether a directed graph $D$ contains a spanning directed tree isomorphic to a given target directed tree $T$. Since the method for solving the case where $D$ contains exactly one weak cycle is similar to that presented in Section 3, we henceforth assume that $D$ contains at least two weak cycles.\newline

\noindent\textbf{Definition 4.1.} Given two anchor points $u, w \in V$, an \textit{anchor chain} between $u$ and $w$ is a sequence of vertices $P=(u,v_1,v_2 ,\cdots,v_m ,w)$ such that for each adjacent pair of vertices $(v_{i-1}, v_i)$ in the sequence (with $v_0=u$ and $v_{m+1}=w$), at least one of the arcs $(v_{i-1}, v_i)$ or $(v_i, v_{i-1})$ exists in the arc set $A$ of the graph.\newline

\noindent\textbf{Lemma 4.1.} Let $F$ be a spanning directed tree in $D$ that is isomorphic to the target tree $T$, and let $G'$ be the graph kernel of $D$. Then, any redundant arc $a \in A_D \setminus A_F$ must lie on an anchor chain that corresponds to an edge in $G'$.

\begin{proof}
Assume, for the sake of contradiction, that exists a redundant arc $a=(u,v)$ that does not lie on any anchor chain. This implies that neither $u$ nor $v$ is an anchor point, meaning they were not preserved in the graph kernel $G'$, nor were they part of a path that was contracted into a single edge of the kernel. Consequently, $u$ and $v$ must be vertices that were removed during a leaf-pruning phase of the kernelization. In that case, the arc $(u,v)$ would serve as a bridge, uniquely connecting two otherwise disconnected components, $C(u)$ and $C(v)$. The removal of this arc would disconnect $F$, which contradicts the premise that $a$ is a redundant arc. Therefore, any redundant arc $a \in A_D \setminus A_F$ must lie on an anchor chain that corresponds to an edge in $G'$. 
\end{proof}

\noindent\textbf{Lemma 4.2.} Suppose the directed graph $D$ contains a directed spanning tree $F$ that is isomorphic to a target tree $T$. If there are $k$ redundant arcs, then there exists a subset of $k$ edges in the graph kernel $G'$, such that each corresponding anchor chain in $D$ contains exactly one redundant arc.

\begin{proof}
Let $A_{R} = A_D \setminus A_F$ denote the set of $k$ redundant arcs. By Lemma 4.1, every arc in $A_{R}$ lies on some anchor chain. Consider an anchor chain $P_{u, w}$ in $D$ corresponding to an edge $(u', w') \in E_{G'}$ of the graph kernel. Removing two or more arcs from $P_{u, w}$ disconnects the chain, partitioning its vertices into at least two connected components, which means at least one component becomes unreachable from other vertices in the graph, violating the spanning tree's connectivity. Consequently, each anchor chain contains at most one redundant arc. With $k$ redundant arcs in total, the pigeonhole principle implies exactly $k$ anchor chains exist, each containing one redundant arc. These $k$ chains correspond to a unique set of $k$ edges in the graph kernel $G'$.
\end{proof}

\noindent\textbf{Theorem 4.1.} Let $P_{u,w}$ be an anchor chain in $D$ corresponding to an edge in the graph kernel $G'$. If $D$ contains a directed spanning tree $F$ rooted at $r_D$, the redundant arc associated with $P_{u,w}$ can be uniquely identified based on the direction of the arcs within the chain.

\begin{proof}
A necessary condition is that the potential root $r_D$ can reach all other vertices in $D$. Note that in the chain $P_{u,w}$, every internal vertex $v \neq r_D$ has in-degree 1, while $r_D$ has in-degree 0 in $F$. Therefore, we categorize cases based on whether $r_D$ lies on $P_{u,w}$.

\noindent\textbf{Case 1: $r_D$ is on $P_{u,w}$ (i.e., $r_D = v_i$ for some $i \in \{1, \dots, m\}$)} If the arcs on the anchor chain have a consistent direction. To satisfy the root condition in-degree($r_D$) = 0, any arc $(v_{i-1}, v_i)$ or $(v_{i+1}, v_i)$ on the chain must be the redundant arc. If there is exactly one arc-direction reversal, each non-root $v_j$ must retain its unique incoming arc. Removing any internal arc $(v_{j-1}, v_j)$ or $(v_j, v_{j+1})$ would disconnect ${v_1,\dots,v_{j-1}}$ (if $j \leq i$) or ${v_{j+1},\dots,v_m}$ (if $j \geq i$) from $r_D$, violating connectivity. Thus, only an end arc ($(u, v_1)$ or $(v_m, w)$) can be redundant arc. If there are two reversals, one reversal occurs at $r_D$, and the other at some $v_j$ with two incoming edges. To satisfy in-degree($v_j$) = 1, one of these two arcs to $v_j$ must be redundant arc.

\noindent\textbf{Case 2: $r_D$ is not on $P_{u,w}$} If the arcs on the anchor chain have a consistent direction, each internal vertex $v_i$ must retain its unique incoming edge, removing any arc $(v_{i-1}, v_i)$ or $(v_{i+1}, v_i)$ will cause $v_i, \dots, w$ or $v_i, \dots, u$ to be unreachable from the root, violating connectivity. Thus, only $(v_m, w)$ or $(v_1, u)$ can be removed. If there is exactly one arc-direction reversal, $v_i$ has two incoming arcs. To satisfy in-degree($v_i$) = 1, one of these two arcs to $v_i$ must be redundant arc.
\end{proof}

The problem, as previously stated, is to find a spanning directed tree within a given directed graph $D$ that is isomorphic to a target directed tree $T$. The algorithm is outlined as follows. First, the algorithm computes the graph kernel $G'$ of the underlying undirected graph of $D$ using the Make-Contractible procedure. Subsequently, it iterates through each vertex $v \in V_D$ as a potential root $r_D$ for the desired spanning tree. For each such potential root, we first verify its reachability to all other vertices in $D$. If this reachability condition holds, the algorithm select $k$ edges from $E_{G'}$ in all $\binom{|E_{G'}|}{k}$ ways. For each such subset of $k$ kernel edges, we identify the corresponding anchor chains in $D$. Then, guided by Theorem 4.1, we uniquely determine and remove one redundant arc from each of these $k$ chains, yielding a candidate spanning tree, $F_{cand}$. This candidate tree $F_{cand}$ is then tested for isomorphism against $T$. A successful match immediately returns YES. If all potential roots and all $k$-edge subsets of the kernel are exhausted without finding an isomorphic tree, the algorithm returns NO. The full procedure is detailed in Algorithm 2.\newline

\begin{figure}[!t] 
    \centering 
    \begin{algorithm}[H] 
      \footnotesize
      \caption{ALG-PDSTIP}
      \label{alg:directed_search_optimized}
      \begin{algorithmic}[1]
        \Statex \textbf{Input:} A directed graph $D=(V_D, A_D)$, a directed tree $T=(V_T, E_T)$, and a parameter $k$
        \Statex \textbf{Output:} YES if a spanning directed tree of $G$ is isomorphic to $T$; and NO otherwise
        
        \State $G \gets \text{underlying undirected graph of } D$
        \State $G' \gets \text{Make-Contractible}(G)$
        \State Map each edge $e' \in E_{G'}$ to its anchor chain $P_{e'}$ in $D$
        
        \ForAll{edge $e' \in E(G')$}
            \State $P_{e'} \gets \text{anchor chain for } e'$
            \State $S[e'] \gets \text{Get Redundant Arcs of $P_{e'}$ by Theorem 4.1. Case 2}$
        \EndFor
        
        \For{each potential root $r_D \in V_D$}
          \If{$\text{not } \text{Reachable}(D, r_D)$}
            \State \textbf{continue}
          \EndIf
          
          \ForAll{edge subsets $E'_{\text{cand}} \subseteq E_{G'}$ of size $k$}
            \State $E_{\text{R}} \gets \emptyset$
            \ForAll{edge $e' \in E'_{\text{cand}}$}
              \State $P_{e'} \gets \text{anchor chain for } e'$
              \If{$r_D \in \text{Vertices}(P_{e'})$}
                \State $E_P \gets \text{Get Redundant Arcs of $P_{e'}$ by Theorem 4.1. Case 1}$
              \Else
                \State $E_P \gets S[e']$ 
              \EndIf
              \State $E_{\text{R}} \gets E_{\text{R}} \cup E_P$
            \EndFor
            
            \ForAll{combinations $\left(e_{\text{rem}}^1, \dots, e_{\text{rem}}^k\right)$ s.t. $e_{\text{rem}}^i \in C_{e'}^i \text{ for } i=1,\dots,k$} 
              \State $E_{\text{del}} \gets \{ e_{\text{rem}}^1, \dots, e_{\text{rem}}^k \}$ 
              \State $F_{cand} \gets (V_D, A_D \setminus E_{\text{del}})$
              \If{$F_{cand}$ is a spanning directed tree and $F_{cand} \cong T$} 
                \State \Return YES
              \EndIf
            \EndFor
          \EndFor
        \EndFor
        
        \State \Return NO
      \end{algorithmic}
    \end{algorithm}
    \captionsetup{skip=0.25pt}
    \caption{The algorithm for the PDSTIP}
    \label{fig:pdstip_algorithm} 
\end{figure}

\noindent\textbf{Theorem 4.2.} Algorithm ALG-PDSTIP can solve the PDSTIP in runtime $O(n^2 \cdot 2^{4k-3})$.

\begin{proof} Given an instance of PDSTIP ($D$, $T$, $k$). ALG-PDSTIP involves converting $D$ to its underlying graph $G$, generating the graph kernel $G'$, and establishing the mapping from each kernel edge back to its corresponding anchor chain in $D$ in step 2 and step 3, which can be completed in linear time $O(n+k)$. Then, the algorithm preprocess redundant arcs of each anchor chains in step 4 to 6, which costs $O(n+k)$ as well. After this, it iterates through each of the $n$ vertices in $V_D$ as a potential root in step 7 to 8. For each of them, a reachability check (e.g., using DFS or BFS) costs $O(n+k)$. Since this must be done for all $n$ potential roots, this phase has a total time complexity of $O(n(n+k))$. By Lemma 3.3 and the edge constraint $|A_D| = |V_D| + k - 1$, the graph kernel $G'$ contains at most $3k-3$ edges. For each potential root, the algorithm enumerates all $k$-edge subsets of $E_{G'}$, generating the following number of combinations in step 10:
$$
\begin{aligned}
     &\binom{3k-3}{k} \leq \sum_{i=0}^{3k-3} \binom{3k-3}{i} = 2^{3k-3}
\end{aligned}
$$
For each of the $k$ chosen kernel edges, we must identify and remove one redundant arc from its corresponding anchor chain. From Step 10 to 18, we can obtain the set of redundant arcs corresponding to the currently enumerated edge subset in linear time $O(k)$. As established in Theorem 4.1, certain arc directions on a chain can create ambiguity, leading to at most two choices for the removable arc (in step 19 to 20). In the worst-case scenario, each of the $k$ chains exhibits such ambiguity, introducing a branching factor of $2^k$. Therefore, for a single potential root, the total number of candidate spanning tree to check is bounded by $O(\binom{3k-3}{k} \cdot 2^k)$. Each of generated candidate is tested for isomorphism against $T$ which takes $O(n)$ time in step 21 to 22. In summary, we can get the total time complexity:
$$
\begin{aligned}
T(n, k) = & \ O\left(n(n+k) + n \cdot \binom{3k-3}{k} \cdot 2^k \cdot (n+k)\right) \\
  =  & \ O\left(n^2 \cdot \binom{3k-3}{k} \cdot 2^k\right) \\
  = & \ O\left(n^2 \cdot 2^{3k-3} \cdot 2^k\right) \\
  =  & \ O(n^2 \cdot 2^{4k-3})
\end{aligned}
$$
\end{proof}

\section{Conclusion}

In this paper, we study the Parameterized Spanning Tree Isomorphism problem and present two fixed‑parameter tractable algorithms---one for the undirected case and one for the directed case. For the undirected version, we develop a FPT algorithm based on a structural reduction of the input graph to a graph kernel of size $O(k)$. We then devise a two-phase matching strategy, resulting in an overall runtime of $O(n^2logn \cdot 2^{k \log k})$. For the directed version, we introduce the concept of anchor chains, proving that all redundant arcs must lie on these chains, which correspond to edges in the graph kernel. This insight allows the problem to be constrained effectively. We propose a deterministic FPT algorithm that solves the directed spanning tree isomorphism problem in $O(n^2 \cdot 2^{4k-3})$ time by enumerating the $\binom{3k-3}{k}$ edge subsets of the kernel and systematically resolving at most $2^k$ local ambiguities arising from arc directions. Our algorithms significantly advance the state of the art for the Spanning Tree Isomorphism Problem and offer novel perspectives for other subgraph isomorphism problems. Future improvements in performance may be achieved through an even more granular analysis of the interplay between the graph and tree structures.

\clearpage

\printbibliography

\end{document}